\RequirePackage{fix-cm}
\documentclass[]{interact}

\usepackage{epstopdf}
\usepackage[caption=false]{subfig}
\usepackage{booktabs}
\usepackage{rotating}
\usepackage{graphicx}
\graphicspath{{figures/}}
\usepackage{url}
\usepackage[numbers,sort&compress]{natbib}
\bibpunct[, ]{[}{]}{,}{n}{,}{,}
\renewcommand\bibfont{\fontsize{10}{12}\selectfont}
\usepackage{hyperref}
\usepackage{xcolor}

\newif\ifshowrev\showrevtrue
\ifshowrev

\else

\fi
\usepackage[nomarkers,nolists]{endfloat}

\begin{document}

\articletype{ARTICLE}

\title{Building a research-software catalog with a coding agent:
from hackathon prototype to public deployment}

\author{
\name{Kazuyoshi Yoshimi$^{a}$\thanks{CONTACT Kazuyoshi Yoshimi. Email: k-yoshimi@issp.u-tokyo.ac.jp}, Satoshi Terasaki$^{a}$, Gotai Yamada$^{a}$}
\affil{$^{a}$Institute for Solid State Physics, The University of Tokyo, Kashiwa, Japan}
}

\maketitle

\begin{abstract}
Generative AI and coding agents can accelerate research software development, but they also increase the need for efficient software discovery and maintenance. We developed a repository catalog during a three-day hackathon and subsequently examined the engineering required to make it suitable for public deployment, including adversarial review, data-quality checks, browser-level validation, and publication safeguards. We then explored whether the lessons learned from this prototype could be transferred to a much larger, human-curated portal, through a retrieval agent under development for MateriApps that combines curated portal metadata, external documentation, vector search, and local language-model generation. Implementation with coding agents was rapid, but achieving reliable operation required substantial additional engineering: the most consequential problems were not crashes but silent failures that produced plausible yet incomplete or incorrect outputs, arising from incomplete data acquisition, misleading assessments, and retrieval or preprocessing failures. These observations suggest that AI-assisted software portals require explicit validation, monitoring, and repeated review, and that curated metadata and maintained documentation remain essential. The MateriApps work is exploratory and remains under active development, so the observations reported for it are preliminary; a comparable combination of curated metadata, automatically collected documentation, and retrieval-based assistance may nevertheless be useful for extending other research-software portals.
\end{abstract}

\begin{graphicabstract}
\includegraphics[width=\textwidth]{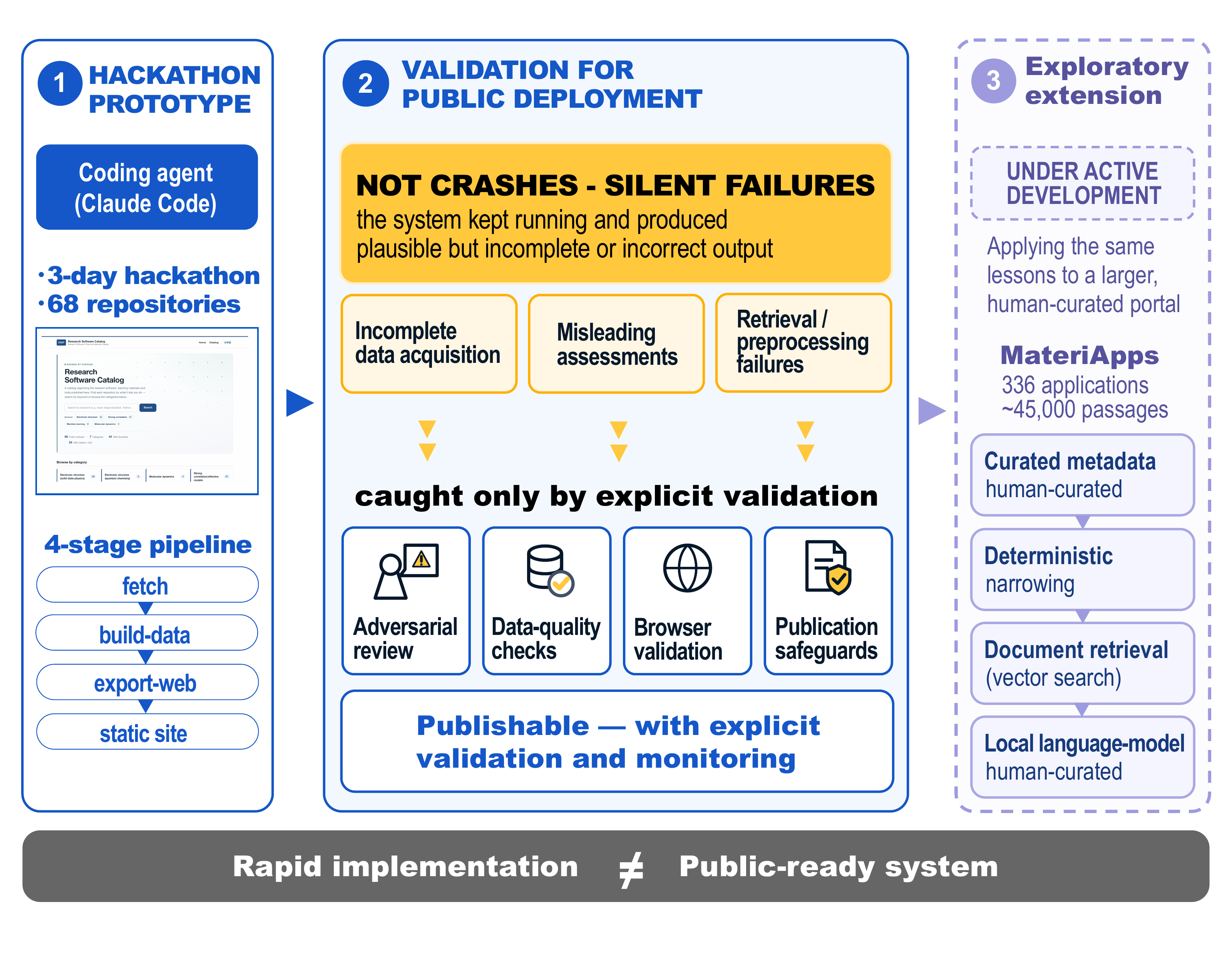}
\end{graphicabstract}


%%%%%%%%%%%%%%%%%%%%%%%%%%%%%%%%%%%%%%%%%%%%%%%%%%%%%%%%%%%%%%%%%%%%%%%%%%%%%%
\section{Introduction}
%%%%%%%%%%%%%%%%%%%%%%%%%%%%%%%%%%%%%%%%%%%%%%%%%%%%%%%%%%%%%%%%%%%%%%%%%%%%%%

Generative AI and coding agents are lowering the cost of developing research software. Since large language models were shown to generate working code from natural-language descriptions\cite{Chen2021}, AI coding assistants such as GitHub Copilot\cite{copilot} have been integrated into everyday software-development workflows, and tools and extensions that previously required substantial programming effort can now be produced within short development cycles. Empirical studies of these tools report genuine gains in development speed alongside recurring defects in the generated code, including quality and security weaknesses that are not evident from the output alone\cite{Yetistiren2023,Pearce2022,Zhang2023}. This acceleration, however, creates a corresponding challenge: software may be developed and updated faster than information about it can be collected, maintained, and made discoverable to potential users.

This problem is particularly relevant to continuing software development programs. The Project for Advancement of Software Usability in Materials Science (PASUMS)\cite{Yoshimi31122025}, for example, produces new software packages and functional enhancements each year. MateriApps\cite{materiapps,Konishi15}, maintained by the Institute for Solid State Physics, provides a community portal through which such computational materials science software can be discovered. As of 31 July 2026 it listed 336 applications across 11 categories. Its descriptions, classifications, and evaluations have largely been created and maintained through human editorial effort. As the number of software packages and updates increases, however, manually keeping this information current becomes increasingly difficult.

The challenge is therefore not simply to create a software portal, but to maintain one as the underlying software ecosystem evolves. Findability is the first of the FAIR principles as they have been adapted to research software\cite{Barker2022}, and it depends on descriptive metadata of the kind that software citation practice also relies upon\cite{Smith2016}. Registries and archives address parts of this need in other communities: bio.tools documents bioinformatics resources through sustained community curation\cite{Ison2016}, while Software Heritage preserves and identifies source code itself\cite{DiCosmo2017}. In all such systems, however, new applications must be registered, and existing entries must be updated when repositories, documentation, supported environments, or functionality change. If software development accelerates while curation remains predominantly manual, the gap between available software and discoverable software will continue to widen.

Generative AI contributes to this problem by accelerating software production, but it may also provide part of the solution. In data-driven materials science, its use in research and education has begun to be examined systematically, including the tasks for which current models are and are not dependable\cite{Misawa2025}. Software discovery and portal maintenance are a natural extension of that question: repository metadata and documentation can potentially be collected, summarized, classified, and searched automatically. This study therefore asks whether the technologies that accelerate research software development can also reduce the recurring effort required to maintain the portals through which that software is disseminated.

This paper presents an experience report originating from a three-day hackathon held in June 2026\cite{aimhack2026}, whose theme was the construction of materials databases using AI agents. During the hackathon, we developed a catalog site that automatically collected information from research software repositories and provided a natural-language interface using a locally hosted language model. The initial system contained 68 catalog records. No further repositories were made public between the hackathon and the 1 September 2026 snapshot analyzed here, so the same 68 records are reported throughout this paper. We then examined the additional engineering and validation required to transform this prototype into a system that could be deployed and maintained for use by others.

After the hackathon and the subsequent development of the catalog, we began exploring whether the lessons learned from this small prototype could be transferred to an established portal of much larger scale. MateriApps was selected as the test case because it is human-curated and because its documentation corpus is far larger and more heterogeneous than that of the catalog. Information collected from the portal and from the external documentation linked by its entries yielded approximately 49,500 indexed passages associated with the 336 applications listed on 31 July 2026, in the corpus examined here, whose crawl was completed on 3 August 2026. To retrieve information from this collection, we assembled a staged workflow combining manually curated portal metadata, deterministic narrowing, document retrieval, and language-model-based response generation. This retrieval agent under development for MateriApps is exploratory and remains under active development; we therefore present it as a larger-scale case study rather than as a second completed system.

The two settings address complementary aspects of portal maintenance. The repository catalog supports the automatic incorporation of newly developed and updated software, including outcomes generated through continuing projects such as PASUMS. The MateriApps case study indicates how automation can build upon, rather than replace, existing editorial work: manually assigned categories and tags remain important for narrowing broad user queries to relevant groups of applications.

This study makes three main contributions. First, it documents the transition from a generative-AI-assisted hackathon prototype to a deployable research software portal. Second, it identifies failure modes in which automated processing produces plausible but incorrect results without explicit execution errors, a class of behavior related to but distinct from the hallucination phenomena documented for language generation itself\cite{Ji2023}. Third, it illustrates how automated collection, human-curated metadata, document retrieval, and language-model generation can play complementary roles in software discovery.

The remainder of this paper is organized as follows. Section~\ref{sec:built} describes the catalog developed during the hackathon, and Section~\ref{sec:publishable} discusses the engineering and validation required to make it suitable for public deployment. Section~\ref{sec:scaling} describes the exploratory application of these lessons to MateriApps. Section~\ref{sec:discussion} discusses the implications for software portal maintenance, system reliability, and human--AI collaboration in software curation, together with the limitations of the study, and Section~\ref{sec:conclusion} concludes.

%%%%%%%%%%%%%%%%%%%%%%%%%%%%%%%%%%%%%%%%%%%%%%%%%%%%%%%%%%%%%%%%%%%%%%%%%%%%%%
\section{What we built first: the catalog}\label{sec:built}
%%%%%%%%%%%%%%%%%%%%%%%%%%%%%%%%%%%%%%%%%%%%%%%%%%%%%%%%%%%%%%%%%%%%%%%%%%%%%%

\subsection{Catalog pipeline and readiness assessment}\label{sec:pipeline}

\begin{figure}
\centering
\includegraphics[width=\textwidth]{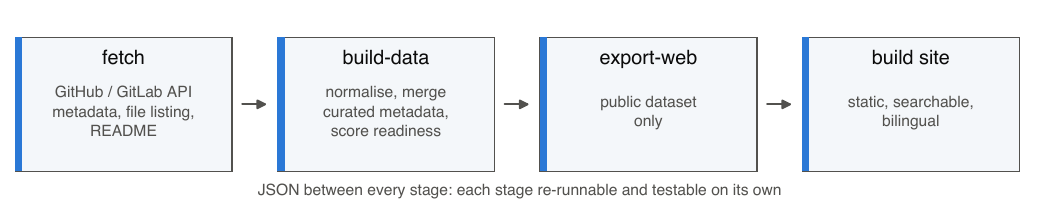}
\caption{The catalog-generation pipeline. JSON interfaces allow the output of
each stage to be inspected, tested, and regenerated independently. They also
allow the last known-good output of an earlier stage to be retained when a later
stage fails, as discussed in Section~\ref{sec:data-honest}.}
\label{fig:pipeline}
\end{figure}

The catalog is generated through a four-stage pipeline
(Figure~\ref{fig:pipeline}). JSON files are used as the interfaces between
stages, allowing each stage to be executed, inspected, and tested independently.

The \textbf{fetch} stage collects repository and project metadata through the
GitHub and GitLab APIs. It handles pagination and rate limits, and records basic
repository metadata, the root-level file listing, and the README content. The
\textbf{build-data} stage normalizes these records, merges manually maintained
metadata---including taglines, categories, intended users, use cases, and
DOIs---applies the publication policy, and computes a readiness score. The
\textbf{export-web} stage extracts only the records approved for public release.
Finally, a static-site generator produces a searchable and filterable website
that can be deployed through GitHub Pages, GitLab Pages, or another static
hosting service. The pipeline is executed on a daily schedule by a GitHub
Actions workflow, so that repository metadata, readiness scores, and the
published site are regenerated without manual intervention; the same workflow
can also be triggered manually. Each scheduled run rebuilds the catalog from
the provider APIs and redeploys the site only after the publication checks
described in Section~\ref{sec:data-honest} have passed. The figures in this
paper are taken from one deployment of the resulting
catalog\cite{repocatalog}.
The same implementation generated three site profiles from separate
configuration files at the revision described here. Repository sources, publication policies, branding, and
presentation settings are specified per deployment rather than embedded in the
application code. This separation improved reuse, although it also introduced
configuration-related security concerns discussed in
Section~\ref{sec:site-honest}.

One of these deployments is the catalog for the Project for Advancement of
Software Usability in Materials Science (PASUMS)\cite{Yoshimi31122025}, a continuing
software-development and enhancement project at the Institute for Solid State
Physics (ISSP)\cite{pasumscatalog}. The harvesting, readiness assessment,
publication filtering, and site-generation mechanisms are shared across the
deployments; the principal differences are the configuration, repository source,
and publication policy. The category taxonomy is an exception, because the
current taxonomy reflects the activities and terminology of ISSP and would need
to be adapted for use by another institution. The PASUMS deployment therefore
provides a practical case for examining whether a repository catalog can
reduce the recurring effort required to maintain an institutional software
portal.

\begin{figure}
\centering
\subfloat[Landing page with browse-by-purpose entry points, category counts, and
a keyword search interface.]{%
\includegraphics[width=0.485\textwidth]{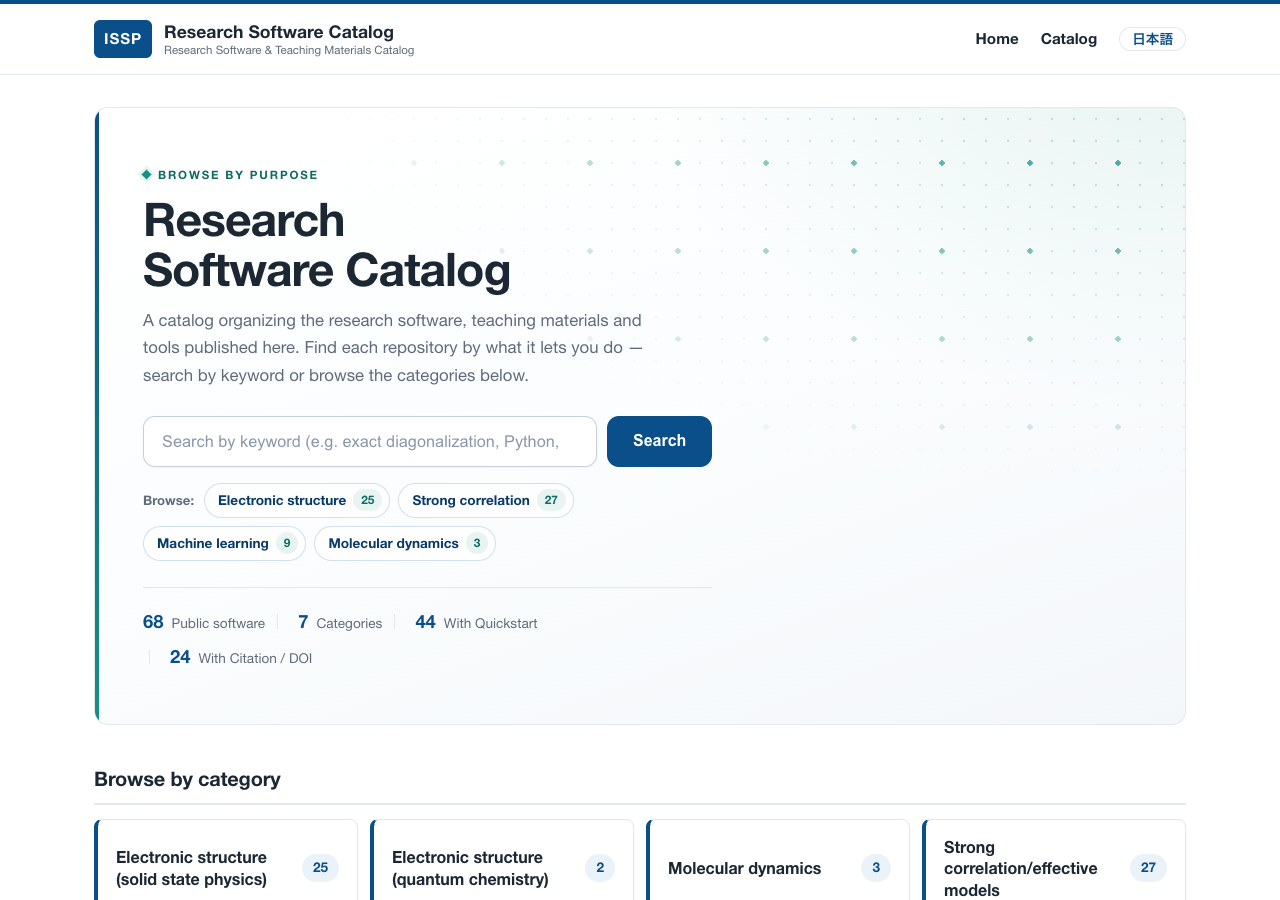}}
\hfill
\subfloat[Catalog listing with keyword search and multi-criteria filtering,
including filters based on missing repository artifacts.]{%
\includegraphics[width=0.485\textwidth]{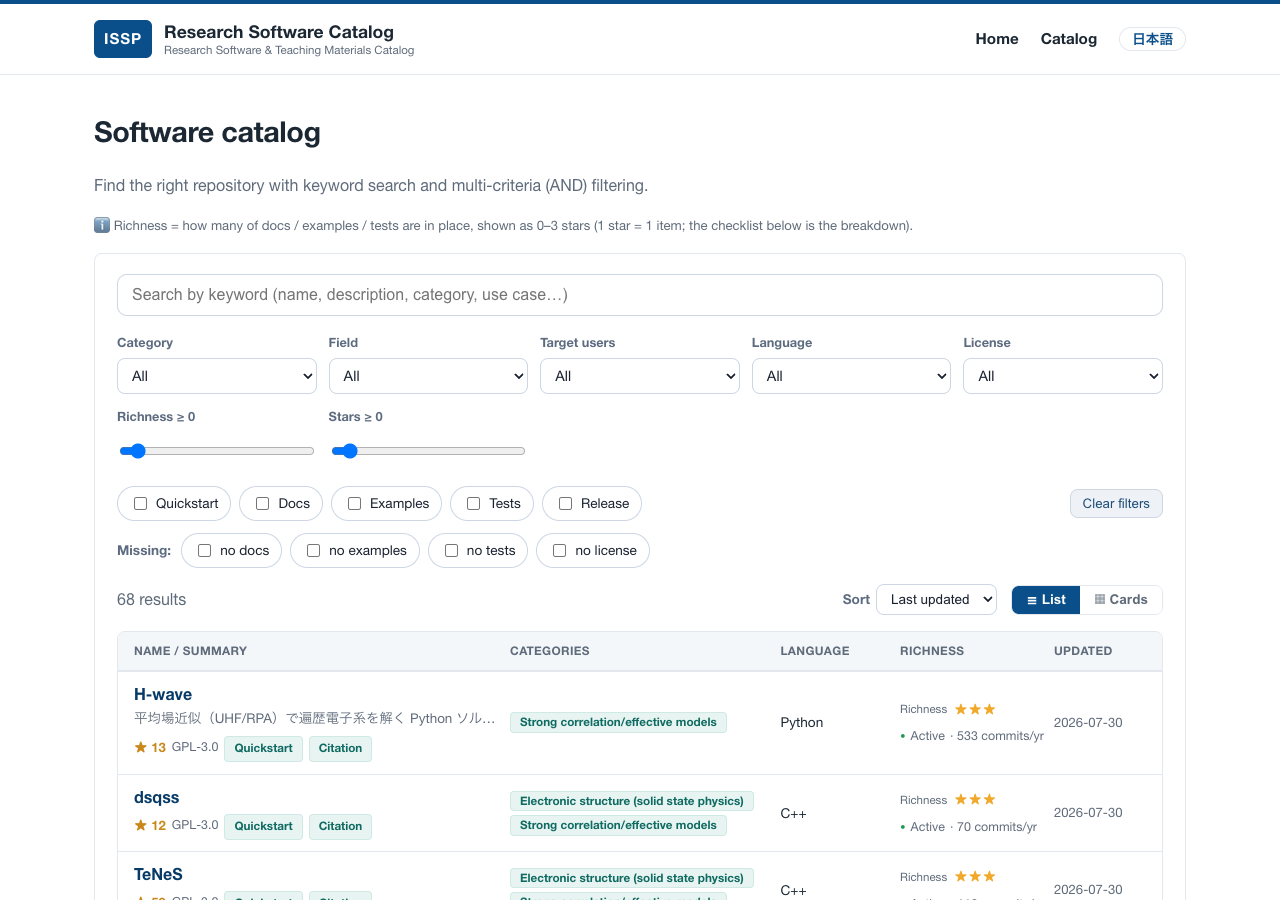}}
\caption{The generated catalog site\cite{repocatalog}. The interface is
available in both English and Japanese. The counts in panel~(a) are calculated
from the same 68 entries analyzed in Figure~\ref{fig:readiness}. Two
additional profiles, including the PASUMS catalog\cite{pasumscatalog}, are
generated by the same implementation using different configurations, repository
sources, and publication policies.}
\label{fig:site}
\end{figure}

The pipeline output is published as a browsable catalog site, available in
both English and Japanese, with browse-by-purpose entry points, category counts,
and keyword search (Figure~\ref{fig:site}).

\begin{figure}
\centering
\includegraphics[width=0.86\textwidth]{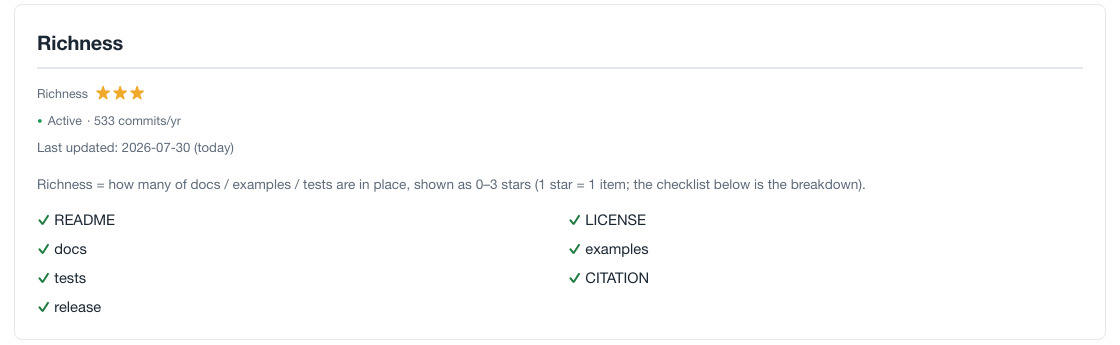}
\caption{Readiness information shown for a repository containing all seven
artifacts. The checklist reports all seven readiness signals, whereas the
Richness rating shown here awards one star for each of documentation, examples,
and tests; the remaining four signals contribute to the separate Openness
rating. Repository activity is displayed separately and is not included in the
readiness score.}
\label{fig:checklist}
\end{figure}

Each repository is assigned a readiness score from 0 to 100 based on seven
binary signals: the presence of a README (20 points), license information
(20), examples (15), tests (15), documentation (10), citation information
(10), and a tagged release (10). The score is presented together with a
checklist (Figure~\ref{fig:checklist}) so that maintainers can identify which
artifacts are missing rather than seeing only an aggregate value.

\begin{figure}
\centering
\includegraphics[width=\textwidth]{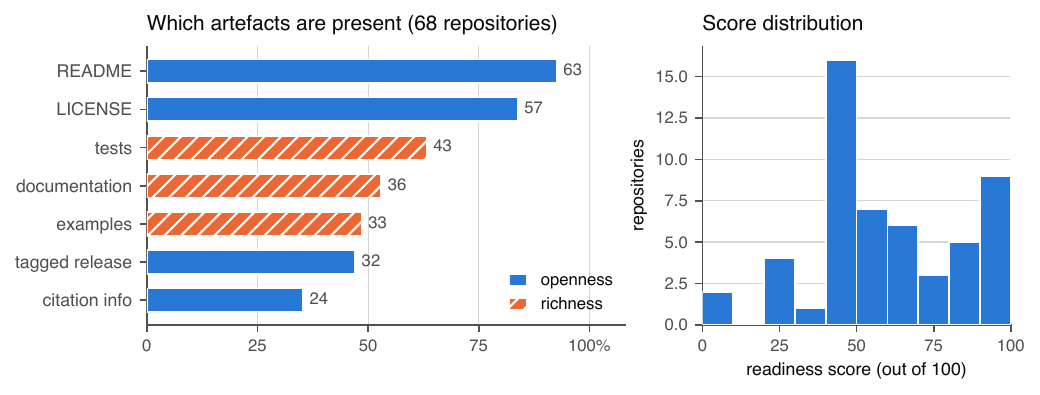}
\caption{Availability of documentation and release artifacts across the 68
catalog entries public on 1 September 2026 and shown in Figure~\ref{fig:site}. README files and
licenses are common, whereas examples, citation information, and tagged releases
are less consistently available. The information that an automated assistant can
provide is necessarily limited by the artifacts exposed by each repository, an
issue revisited in Section~\ref{sec:ceiling}.}
\label{fig:readiness}
\end{figure}

The same signals are also summarized as two star ratings, \emph{Openness} and
\emph{Richness}. Openness is derived from the README, license, citation
information, and release status, and reflects whether an external user can
identify, use, and cite the software. Richness awards one star for each of
documentation, examples, and tests, and is therefore shown on a 0--3 scale
(Figure~\ref{fig:checklist}); it reflects whether the software is explained,
demonstrated, and checked. These indicators are intended as transparent summaries of observable
repository artifacts rather than as evaluations of scientific quality. Across
the 68 entries public on 1 September 2026, these artifacts are unevenly available: README files and
licenses are almost universal, whereas examples, citation information, and
tagged releases are not (Figure~\ref{fig:readiness}).

\subsection{Single-stage question answering}\label{sec:ask}
The initial catalog also includes a lightweight question-answering function implemented as an exploratory feature rather than as part of the published service. It is deliberately disabled in public deployments: the public-build mode removes the feature, and a deployment check prevents publication of any build in which it remains reachable. The function therefore operates only in a private local environment using a locally hosted language model.

A query is tokenized into Latin-script word sequences and, for Japanese and Chinese text, overlapping character bigrams. Catalog records are ranked solely by weighted token overlap across their fields; the language model is used only after retrieval, with the highest-ranked records supplied as context through Ollama\cite{ollama}. The weights reflect the catalog's primary use case of software discovery. Short, structured fields receive greater weight than the README body, with manually curated fields such as the display name, tagline, and use cases assigned the highest weights. Repository names and primary programming languages also receive relatively high weights, whereas the README has the lowest weight among the twelve indexed fields. This reduces the influence of long-form repository text relative to concise metadata prepared specifically for software discovery.

Retrieval is restricted to the public dataset, so internal repository information is never included in the context passed to the model. A Python implementation supports command-line use, while a TypeScript implementation performs retrieval in the browser. In both cases, the intended configuration uses a locally hosted model so that neither queries nor retrieved context need to be sent to an external service.

This first implementation uses single-stage token-overlap retrieval, without query decomposition, learned re-ranking, or answer verification. Its limitations became apparent even over the 68 public records. For example, a query for variational Monte Carlo applications also returned software implementing ordinary Monte Carlo methods, because token overlap alone cannot reliably distinguish a specific method from the broader family to which it belongs. This was acceptable for an exploratory feature that was not exposed to external users. Nevertheless, it made the underlying limitation clear: because retrieval-augmented generation relies on retrieved documents as context for generation\cite{Lewis2020,Gao2023}, retrieval quality becomes critical when questions concern specific software capabilities rather than named software. Section~\ref{sec:scaling} therefore examines how the same limitation reappeared when a related approach was applied to a substantially larger corpus. Before that, however, Section~\ref{sec:publishable} describes the engineering and review required to make the catalog itself suitable for public deployment.

%%%%%%%%%%%%%%%%%%%%%%%%%%%%%%%%%%%%%%%%%%%%%%%%%%%%%%%%%%%%%%%%%%%%%%%%%%%%%%
\section{Making the catalog publishable}\label{sec:publishable}
%%%%%%%%%%%%%%%%%%%%%%%%%%%%%%%%%%%%%%%%%%%%%%%%%%%%%%%%%%%%%%%%%%%%%%%%%%%%%%

The catalog was developed using Claude Code\cite{claudecode}, a command-line coding agent. Over the eight days from the first commit to the completion of the initial development and review cycle, the repository accumulated 54 commits. The development history can be broadly divided into two phases. The first was a construction phase focused primarily on feature implementation. The second was an adversarial review phase, during which most commits addressed a numbered issue. In total, 25 issues were filed.

We organize this section around these issues rather than the sequence of features added during construction. The issues provide a more informative view of the work required to transform a functioning prototype into a system suitable for public deployment. They fall into three groups: the development process used with the coding agent, the integrity of the published data, and the reliability and security of the published site.
\subsection{Working with the coding agent}\label{sec:working}

\subsubsection{Separating implementation from adversarial review.}

During the construction phase, Claude Code\cite{claudecode} was used as the primary coding agent to implement new features. At the end of this phase, Claude Code was instructed to invoke Codex\cite{openaicodex} in a fresh context to perform an adversarial review of the repository. Codex was asked to identify potential failure modes and to report each finding using a common structure consisting of a description of the problem, a concrete failure scenario, and a proposed correction. The resulting issues were then addressed individually.

After these corrections had been applied, the repository was subjected to a further adversarial review. This second pass identified additional problems in both the data pipeline and the front end, which were addressed in two further commits. The workflow therefore did not treat review as a single terminal step; instead, implementation, independent inspection, correction, and renewed inspection formed an iterative cycle.

Although this procedure does not establish reviewer independence in a formal sense, separating implementation and review across different agents and fresh conversational contexts helped expose assumptions that had not been examined during construction. In this case, rapid implementation accounted for only part of the agent-assisted development effort, while repeated attempts to identify failure modes constituted a substantial fraction of the total engineering work.

\subsubsection{Verifying claims against the running system.}
Claims that a change had been correctly implemented were verified against the
deployed or locally running application rather than accepted on the basis of
code inspection alone. Commit records therefore include browser-level
verification notes, such as confirming that switching the interface language
preserves the active filters, or that a relative date is calculated against the
current date rather than the site-build date.
The latter exposed a failure mode specific to static-site generation. Relative
timestamps had been calculated at build time and embedded in the generated
pages, causing a continuously deployed site to display increasingly stale
values. The footer also labeled the site-build time as the time at which the
underlying dataset had been updated. Both outputs were syntactically valid and
visually plausible, but represented the wrong quantities.
This experience highlighted a distinction between verifying source-code changes
and verifying observable system behavior. For generated sites in particular,
correctness may depend on when and where a value is computed, not merely on
whether the calculation itself is implemented correctly.

\subsection{Maintaining the integrity of published data}\label{sec:data-honest}

The most consequential issues concerned the catalog data. Because the system
assigns readiness indicators to repositories, it publishes an assessment rather
than merely reproducing source metadata. Incorrect data presented as a valid
assessment can therefore be more misleading than data that are explicitly
marked as unavailable.

\subsubsection{Distinguishing unavailable data from negative evidence.}

Several early failure modes produced plausible but incorrect outputs without
raising exceptions. For example, if a README request failed because of rate
limiting or a server-side error, the repository was recorded as having no README
and received a lower readiness score. This was indistinguishable from a
repository that genuinely lacked a README. The pipeline could therefore
complete successfully and publish an incorrect assessment even when data
acquisition had failed.

Authentication failures created a similar problem: if one data source could not
be accessed, the output could still pass validation as long as records from
other sources remained. A failed run could also overwrite the last known-good
dataset with partial output. In addition, API requests and rate-limit waits were
not properly bounded, allowing an unresponsive external service to occupy a
continuous-integration job for an extended period.

The revised pipeline treats acquisition failures as explicit data-quality
events. Enrichment failures are counted, and in the deployment configuration
used in this study, a strict run fails if at least five requests are attempted and more than
20\% of them fail. This threshold is an operational choice for the present system
rather than a general standard. Strict mode is enabled in continuous
integration, whereas local runs may continue while clearly marking the resulting
data as degraded.

These changes led to three operational principles: acquisition failures should
not be silently converted into negative values; deployment should stop when
data degradation exceeds an explicit threshold; and a failed run should never
overwrite the last known-good dataset.

\subsubsection{Verifying privacy constraints before deployment.}

A central requirement of the system is that information from private repositories must not be exposed through the public catalog. This requirement is checked at deployment through three mechanisms: a leak check over the exported public dataset, a guard against unexpectedly empty output, and an inspection of the generated site. Deployment proceeds only if all checks pass in public-build mode, which excludes local-only routes.

Adversarial review identified cases in which this requirement could be violated even when individual components behaved correctly in isolation. The most instructive was that operations on internal records could indirectly modify records that had already been validated for publication, because the public and internal datasets initially shared the same record objects. Slug deduplication over the internal dataset, for example, could rewrite a slug that had already passed the leak check, so that the presence of a private record changed a public URL.
The review also found that deployment could proceed without running the test suite, and that the example workflow in the README omitted both the leak checks and public-build mode. The workflows were therefore revised so that tests run on every push and pull request and are also required before deployment.
A separate risk was that internal documentation could be added accidentally to the public repository because it was not covered by the version-control exclusion rules. The repository configuration was revised accordingly.

These findings show that privacy requirements should be verified explicitly at the publication boundary rather than assumed from the correctness of individual components. The deployment checks themselves must also be tested and included in the documented publication procedure.

\subsubsection{Validating rule-based indicators against repository data.}

The readiness indicators are based on simple rules inferred from repository structure, metadata, and documentation. During development, several rules that appeared reasonable in isolation produced incorrect results when applied to real repositories. For example, test detection initially relied on the presence of a \texttt{tests/} directory, but was later extended to inspect test configuration files, README sections, and continuous-integration settings.

Inspection of the generated catalog revealed both false positives and false negatives. A project name containing \texttt{-test} could be mistaken for evidence of a test suite, while a C project with a large shell-based test harness could be classified as a shell project. Similar adjustments were required for license and citation detection to account for repository-specific filenames and wording.

These examples show that simple rule-based indicators can be useful, but must be validated against the repositories to which they are applied. Many misclassifications were difficult to identify from the rules alone but became obvious once the results were inspected in the rendered catalog.

\subsection{Maintaining the reliability of the published site}
\label{sec:site-honest}

\subsubsection{Testing the interface in real browsers}

The front end was evaluated not only through code inspection but also against accessibility guidelines\cite{wcag22} and actual browser behavior. This revealed defects that were difficult to identify from the source code alone.
For example, several filter controls were assigned the same identifier because their identifiers were generated from Japanese labels. This caused accessibility problems and, in some cases, made a label activate the wrong input. Browser testing also revealed that frequent URL updates could exceed Safari's history-update limits, while client-side rendering could leave the main catalog page temporarily blank until JavaScript loaded.

These cases show that an implementation that appears correct at the source-code level can still fail when combined with browser-specific constraints and accessibility requirements.

\subsubsection{Avoiding unnecessary data in the browser.}

Another issue concerned the JavaScript bundle sent to the browser. A filter component imported helper functions from a module that also loaded the complete catalog data. As a result, the full catalog, including README text, was unnecessarily included in the client-side bundle.
The helper functions were moved to a separate module with no dependency on the catalog data. This reduced the relevant JavaScript bundle from 455\, kB to 16\, kB.

This case shows that code organization affects not only maintainability but also the amount of data transferred to and exposed in the browser.

\subsubsection{Validating configuration values.}

To support multiple catalog deployments, site-specific settings such as colors and watermark images were moved into configuration files. Because some of these values were inserted into CSS, invalid or unexpected values could affect the generated page in unintended ways.
The revised implementation validates color values and percent-encodes image paths before inserting them into CSS. These checks are implemented as pure functions and covered by automated tests.

This case shows that configuration values should be treated as inputs that require validation rather than as inherently trusted data.

Taken together, these issues show that constructing a functioning prototype was only part of the development process. Reliable publication also required browser-level testing, careful control of the data sent to the client, and validation of configuration values. The most consequential problems were often not crashes, but cases in which the site continued to operate while producing incomplete, misleading, or unsafe results. The following section examines how related challenges appeared when these lessons were applied to the much larger and more heterogeneous MateriApps corpus.

\section{Applying the approach to a larger existing portal}
\label{sec:scaling}

\begin{figure*}[t]
\centering
\includegraphics[width=\textwidth]{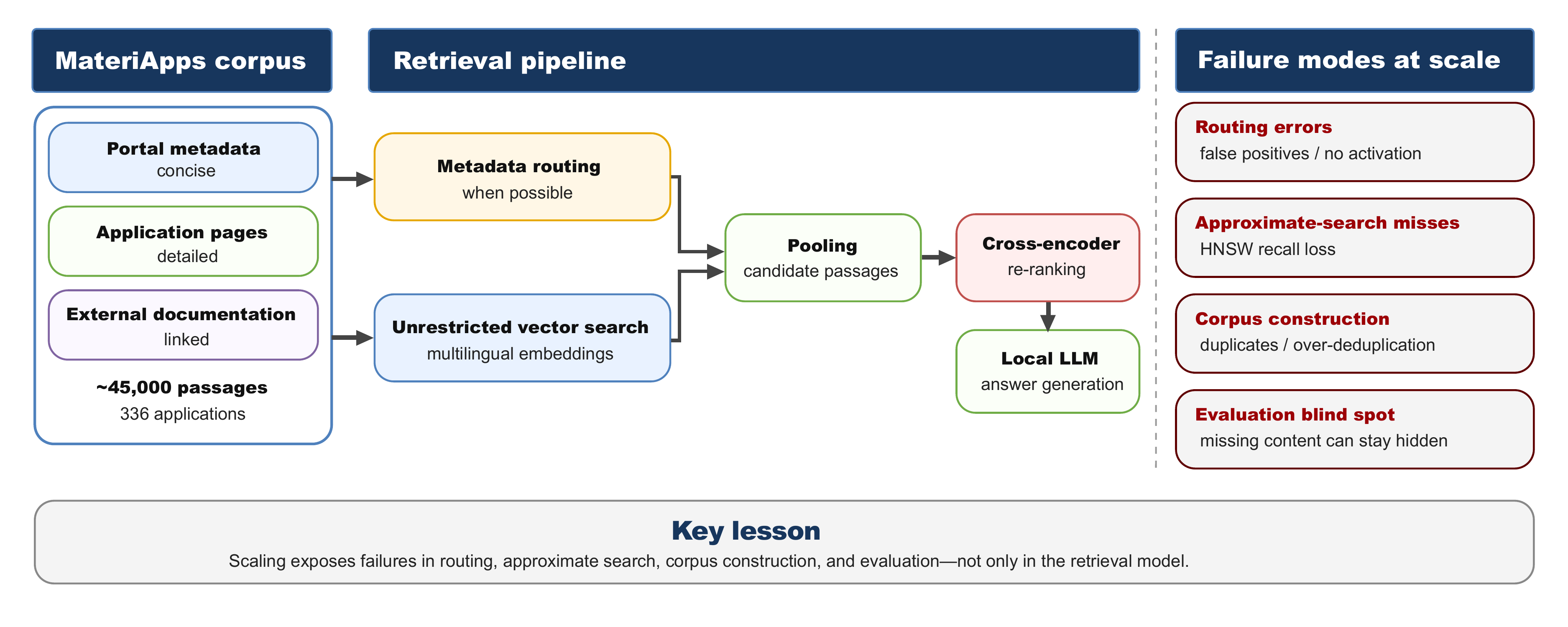}
\caption{Overview of the retrieval workflow applied to MateriApps and the main failure modes exposed at larger scale. Curated metadata routing and unrestricted vector search provide complementary candidate passages, which are pooled and re-ranked by a cross-encoder before answer generation with a locally hosted language model. Scaling revealed limitations in routing, approximate nearest-neighbor search, corpus construction, and evaluation.}
\label{fig:materiapps-retrieval}
\end{figure*}

Following the hackathon and the subsequent development of the repository catalog, we began exploring whether the lessons learned from this small prototype could be transferred to a larger, existing software portal. For this purpose, we applied a related retrieval workflow to MateriApps\cite{materiapps}, a human-curated portal for materials-science software. Including the external documentation linked from its entries, the corpus examined here---its crawl completed on 3 August 2026, covering the 336 applications listed on 31 July 2026, with both the Japanese and English portal pages indexed---comprises 49,485 indexed passages, and is therefore far larger and substantially more heterogeneous than the 68-record catalog of Section~\ref{sec:built}.
This work is still exploratory and remains under active development. Nevertheless, the larger setting has already exposed limitations that did not appear in the initial catalog.
Figure~\ref{fig:materiapps-retrieval} summarizes the retrieval workflow used in the current MateriApps prototype and the main failure modes that emerged as the corpus was scaled up.

\subsection{Retrieval workflow and routing}

The MateriApps corpus was organized into three levels: concise portal metadata, detailed application pages, and external documentation. Retrieval combines these sources in stages.
Curated metadata is first used, when possible, to identify candidate applications. Passages retrieved from the identified candidate applications are then pooled with those returned by unrestricted vector search across all applications, using multilingual sentence embeddings\cite{Wang2024}. A cross-encoder\cite{Nogueira2019}, built on the BGE-M3 multilingual encoder\cite{Chen2024}, then re-ranks the pooled passages before they are passed to a locally hosted language model.

Early experiments suggest that these stages address different types of retrieval failure, but also that the conditions under which each is beneficial are narrower than initially expected. Dictionary routing is effective when the question explicitly contains an application name, but cannot contribute when no such name is detected. On a set of paraphrased keyword questions with the gold answers unchanged, the Japanese dictionary never fired, and routed retrieval was therefore identical to unrestricted vector search. In English, all seven dictionary activations were false positives, causing the routing step to discard documents that unrestricted vector search had retrieved. Cross-encoder re-ranking performed best by mean reciprocal rank, among the modes that do not first restrict the candidate set of applications, on these questions because it re-orders candidates returned by unrestricted vector search rather than committing to a restricted candidate set at an earlier stage.
Table~\ref{tab:materiapps-modes} reports these measurements. The question sets hold the editorially assigned gold applications fixed (34 Japanese and 33 English keyword questions, one per keyword shared by two or more applications) and differ only in wording: the template form embeds the keyword string verbatim, whereas the paraphrased form describes the research situation without it. With the template wording, dictionary routing is near-perfect; with the paraphrase, it is indistinguishable from unrestricted vector search.

\begin{table}
\centering
\caption{Retrieval on keyword questions over the MateriApps corpus (hit@10 / mean reciprocal rank, top-10 passages, questions asked and answered in one language). Template questions contain the keyword string; paraphrased questions keep the same gold applications but avoid it. For the keyword \emph{Ising Model} (six gold applications), for example, the template question is ``Which apps are related to Ising Model?'' and the paraphrased question is ``Which apps can simulate lattices of binary up-or-down spins coupled to their nearest neighbours to study magnetic ordering?''; for \emph{Dynamical Mean Field Theory (DMFT)} (twelve gold applications) the paraphrase is ``For materials where local electron repulsion dominates, which solvers approximate the lattice by one interacting site embedded in a dynamically determined bath?''. The Japanese sets are constructed in the same way. Values are for the routing dictionary after the false-positive fix; before it, English routed retrieval on the paraphrased questions scored 0.515 / 0.252 (seven activations, all on questions that did not concern the matched application). A two-stage hierarchical variant not shown here, which first selects candidate applications by embedding similarity and then searches only within them, reached 0.676 / 0.355 (Japanese) and 0.606 / 0.434 (English) on the paraphrased questions.}
\label{tab:materiapps-modes}
\begin{tabular}{lcccc}
\toprule
 & \multicolumn{2}{c}{Japanese ($n=34$)} & \multicolumn{2}{c}{English ($n=33$)} \\
\cmidrule(lr){2-3}\cmidrule(lr){4-5}
Retrieval mode & template & paraphrased & template & paraphrased \\
\midrule
Unrestricted vector search & 0.912 / 0.700 & 0.647 / 0.332 & 0.970 / 0.700 & 0.667 / 0.291 \\
Dictionary routing & 1.000 / 1.000 & 0.647 / 0.332 & 1.000 / 1.000 & 0.667 / 0.291 \\
Cross-encoder re-ranking & 1.000 / 0.880 & 0.618 / 0.457 & 0.939 / 0.889 & 0.636 / 0.417 \\
Routing $+$ re-ranking (workflow) & 1.000 / 0.914 & 0.618 / 0.457 & 1.000 / 0.985 & 0.636 / 0.417 \\
\bottomrule
\end{tabular}
\end{table}

\subsection{Scaling of approximate retrieval}

The larger corpus has also exposed problems beyond the choice of retrieval model. As the index grew,
the hierarchical navigable small-world index\cite{Malkov2020} used for approximate nearest-neighbor search failed to retrieve some passages that were nearby in embedding space when used with its default search setting. Increasing the parameter controlling the breadth of graph traversal recovered much of the lost performance.
The effect was first observed when the Japanese-only index grew from about 20,000 to about 73,000 passages, where top-5 hit rate for unrestricted vector search fell from 1.000 to 0.950 and returned to 0.990 once the search-breadth parameter (\texttt{ef\_search}) was raised from the library default of 100 to 500. Table~\ref{tab:materiapps-hnsw} repeats the comparison on the cleaned bilingual index examined in this paper; the loss at the default setting persists and is concentrated in questions about individual applications, both factual ones (license, developer, environment) and usage-related ones.

\begin{table}
\centering
\caption{Unrestricted vector search on the 49,485-passage MateriApps index at the two HNSW search-breadth settings. Template questions of eight kinds, up to 25 per kind (197 Japanese, 200 English); the value 500 is the setting used throughout this paper.}
\label{tab:materiapps-hnsw}
\begin{tabular}{lcccccc}
\toprule
 & \multicolumn{3}{c}{Japanese ($n=197$)} & \multicolumn{3}{c}{English ($n=200$)} \\
\cmidrule(lr){2-4}\cmidrule(lr){5-7}
\texttt{ef\_search} & hit@5 & hit@10 & MRR & hit@5 & hit@10 & MRR \\
\midrule
100 (library default) & 0.914 & 0.924 & 0.815 & 0.880 & 0.885 & 0.798 \\
500 & 0.949 & 0.959 & 0.849 & 0.920 & 0.930 & 0.836 \\
\bottomrule
\end{tabular}
\end{table}

\subsection{Corpus construction and hidden failure modes}

Corpus construction introduced additional difficulties: a crawler could collect disproportionate amounts of repeated material from a single site, while aggressive deduplication could remove valid documentation together with boilerplate.

One such failure was particularly instructive.
Two packages had been registered twice in the portal, causing the crawler to collect their official documentation once under each registration. A cleaning rule then discarded any passage occurring under more than one registration, on the assumption that text shared between applications was boilerplate rather than documentation specific to a single application. Because the rule removed every copy rather than retaining one, the documentation of such a package disappeared entirely. For one of the two, the 22 installation pages present in the crawled data were reduced to none in the index, leaving only 17 passages, compared with 183 produced by the corrected pipeline. Aggregate benchmark results did not reveal this defect because a usage question was counted as answered whenever any passage from the correct application was retrieved, and the package's surviving README passage was sufficient to satisfy that criterion. The defect became visible only when a real user asked how to install the package: retrieval correctly narrowed the search to that package, but the relevant installation documentation was no longer present in the index.

These observations should therefore be regarded as early findings from the retrieval agent under development for MateriApps, rather than as a final evaluation of the system. They nevertheless reinforce a lesson already visible in the smaller catalog: a system can remain operational and produce plausible results even when important information has been lost. Ongoing development will focus on retrieval quality, validation of corpus construction, monitoring of application-level coverage, and
extending the leakage-controlled evaluation framework from retrieval performance to generation quality, which has so far been measured only for the English paraphrased keyword questions and not yet for Japanese or for usage questions.

\section{Discussion}\label{sec:discussion}

\subsection{The continuing value of editorial curation}

This study began from the concern that manual curation may not keep pace with the growing volume of research software. Our observations do not resolve that broader question, but they suggest that existing editorial metadata can remain useful when combined with automated retrieval.

In the retrieval agent under development for MateriApps, manually assigned application names can be used directly to restrict the search space before vector retrieval, and this step requires neither a language model nor an additional learned classifier. Its benefit, however, is confined to questions in which the curated string appears verbatim; where it does not, the same mechanism contributes nothing over unrestricted retrieval. Curated metadata therefore guides retrieval for users who already know what to call what they are looking for, and the harder case remains open. A similar pattern appears in the repository catalog, where manually prepared fields such as software names, taglines, and use cases receive greater retrieval weight than automatically harvested README text.

These observations suggest that automated retrieval and generative AI are better viewed as extensions of editorial work than as replacements for it. Curated metadata provides a compact representation of domain knowledge that can guide automated systems before they search larger and less structured collections of documentation.

\subsection{Documentation sets a ceiling on automated discovery}
\label{sec:ceiling}

Retrieval quality is ultimately limited by the information available in the source corpus. In the corpus examined here, whose crawl was completed on 3 August 2026, the crawler obtained no external documentation for 62 of the 336 MateriApps applications: one entry lists no official site or manual, and for the remaining 61 no page could be retrieved or retained. For these applications, improvements in retrieval architecture cannot recover information that is absent from the available sources.

The repository catalog points to a related limitation. Its readiness indicators measure artifacts such as a README, license, examples, tests, citation information, documentation, and tagged releases. These are also among the materials both human users and automated systems need to understand what software does, how it should be used, and how it should be cited.

Preparing research software for machine-assisted discovery therefore does not require a fundamentally new documentation practice. Rather, it reinforces established practices for making software understandable and usable: clear descriptions, explicit licensing, reproducible examples, citation guidance, and identifiable releases benefit both human readers and automated systems.

\subsection{From prototype to sustainable operation}

The catalog and the retrieval agent examined here were developed by a small group closely involved in their maintenance. A more demanding test is whether an institutionally maintained service can remain accurate over time without continuous supervision by its original developers.

The PASUMS catalog\cite{pasumscatalog} is an initial step in this direction. It shows that the same generator can be applied to a different repository organization with its own branding, data sources, and publication policy. However, it has not yet operated long enough to determine whether automated generation actually reduces long-term maintenance effort rather than shifting that effort from writing records to reviewing them.

This distinction matters because many of the failures identified in this study were silent: the system continued to operate while producing incomplete or misleading results. This resembles the broader problem of data and configuration failures propagating silently through machine-learning systems\cite{Sculley2015,Sambasivan2021}. Long-running services therefore require automated checks for changes in coverage, failed data acquisition, and other forms of degraded output rather than relying on routine inspection by the original developers.

\subsection{Limitations and future work}

This work is an experience report based on the repository catalog and the retrieval agent under development for MateriApps, both built by a small group of developers. Its conclusions should therefore be interpreted within that scope. In particular, we did not perform a controlled comparison between agent-assisted and conventional software development. We therefore cannot determine which of the observed defects were specific to coding-agent use and which would also arise in conventional development.

The preliminary retrieval evaluation relies on automatically generated questions rather than a human-authored gold standard, and such questions cannot fully represent real user information needs. Using a local language model as a judge\cite{Zheng2023} to assess generation quality introduces further assumptions and possible biases. The real-user query that revealed missing installation documentation further shows that aggregate benchmark performance does not guarantee application-level corpus completeness.

Two extensions are particularly relevant for future work. First, automatically derived repository-readiness indicators could be compared with MateriApps editorial assessments to examine where machine-readable evidence agrees with or differs from expert judgment. Second, the retrieval agent could be connected to executable environments such as MateriApps LIVE!\cite{MAL-paper}, which is one possible route from software discovery to reproducible, human-supervised computational workflows.

While the catalog continues to be maintained, the retrieval agent under development for MateriApps remains exploratory, and the observations reported for it are correspondingly preliminary. The catalog and corpus quantities reported here---the record, application, passage, and documentation counts---describe the two systems as they stood at their respective snapshot dates: the catalog on 1 September 2026, the MateriApps portal listing on 31 July 2026, and the crawled corpus on 3 August 2026. Both continue to change, so these values should be read as a dated snapshot rather than as stable properties of either system. 
The complete per-record export of the 1 September catalog build, together with the derived tables behind Figure~\ref{fig:readiness}, and the MateriApps corpus behind the retrieval case study (portal metadata, crawled documentation pages, and the vector index) are archived in the ISSP Data Repository\cite{cataloguedata} (see Data availability).

%%%%%%%%%%%%%%%%%%%%%%%%%%%%%%%%%%%%%%%%%%%%%%%%%%%%%%%%%%%%%%%%%%%%%%%%%%%%%%
\section{Conclusion}\label{sec:conclusion}
%%%%%%%%%%%%%%%%%%%%%%%%%%%%%%%%%%%%%%%%%%%%%%%%%%%%%%%%%%%%%%%%%%%%%%%%%%%%%%

We investigated whether generative-AI technologies that accelerate research
software development can also support its discovery and maintenance. The central
system in this study is a repository catalog developed during a three-day
hackathon and subsequently hardened for public deployment. We also explored
whether the lessons from this prototype transfer to a larger, human-curated
portal through a retrieval agent under development for MateriApps; this work
remains exploratory, and its observations are preliminary.

Coding agents enabled rapid implementation, but reliable publication required
repeated review and validation. The most consequential failures were not
crashes, but silent failures that produced plausible yet incomplete or incorrect
outputs. Addressing them required separating implementation from adversarial
review, surfacing acquisition and processing failures explicitly, testing
publication safeguards, and checking system behavior against the running
application rather than source code alone.

Our observations also indicate that automated retrieval builds upon rather than
removes the need for editorial curation and maintained documentation. Curated
metadata helped guide retrieval for users who already knew what to call what
they were looking for, while missing documentation could not be recovered by
retrieval architecture alone. The PASUMS deployment provides an
early institutional example, and similar combinations of curated metadata,
automatically collected documentation, and retrieval-based assistance may be
applicable to other research-software and research-data portals such as
MatDaCs\cite{matdacs}. Connecting software discovery to executable environments
such as MateriApps LIVE!\cite{MAL-paper} is one possible further direction.

Taken together, these observations suggest that generative AI is most useful in
research-software infrastructure when it extends, rather than replaces, existing
technical and editorial work. 
This work provides one concrete case study of where generative AI can and cannot be relied upon in research-software infrastructure for data-driven materials science~\cite{Misawa2025}. 
The central challenge is not simply to generate more
software or more answers, but to ensure that software, metadata, documentation,
and automated outputs remain verifiable, maintainable, and useful to the wider
research community.

%%%%%%%%%%%%%%%%%%%%%%%%%%%%%%%%%%%%%%%%%%%%%%%%%%%%%%%%%%%%%%%%%%%%%%%%%%%%%%

\section*{Acknowledgements}
The initial version of the catalog described in this paper was developed at DxMT AIMHack 2026~\cite{aimhack2026}. We thank the organizers and the other participants for discussions during and after the event.
G.Y. used the ARIM-mdx data system~\cite{hanai2024arim} in this work.
This work was supported by the Ministry of Education, Culture, Sports, Science and Technology (MEXT), Japan, through the Data Creation and Utilization-Type Material Research and Development Project (JPJ010337) and the project ``Developing a Research Data Ecosystem for the Promotion of Data-Driven Science''.
K.Y. and S.T. were additionally supported by the Japan Science and Technology Agency (JST) through the Moonshot Research and Development Program, Grant Number JPMJMS24A3, for the exploratory work on applying this approach to a larger research-software portal, reported in Section~\ref{sec:scaling}, and for the preparation of this manuscript.

\section*{Disclosure statement}
No potential conflict of interest was reported by the author(s).

\section*{Data availability}

The generated catalog\cite{repocatalog} and its institutional deployment for
ISSP PASUMS\cite{pasumscatalog} are publicly accessible, as is
MateriApps\cite{materiapps}, the portal for which the retrieval agent is being
developed.
These are live services that continue to change, so they do not by themselves
reproduce the dated snapshots analyzed here.

The data underlying this article are archived in the ISSP Data
Repository\cite{cataloguedata}. The archive contains, first, the complete
per-record export of the 1 September 2026 catalog build analyzed in
Section~\ref{sec:pipeline} (68 repositories, with the seven readiness signals
and the readiness score of each record), the derived signal table and the
aggregate values of Figure~\ref{fig:readiness}, provenance metadata of the
generating build, and a validation script that checks all reported values;
and second, the corpus behind the MateriApps retrieval case study of
Section~\ref{sec:scaling}: the portal metadata of the 336 applications listed
on 31 July 2026, the crawled documentation pages (crawl completed on
3 August 2026), the 49,485-passage vector index built from them, and
supplementary generation-quality evaluation runs. The catalog data and
evaluation runs are released under a CC BY 4.0 license; the crawled
documentation pages contain text from third-party websites and manuals, which
remains under the rights of its owners and is included solely to allow
verification of the reported results.

The source code of the catalog generator and of the retrieval agent under
development for MateriApps, together with the question sets and per-question
retrieval results behind Tables~\ref{tab:materiapps-modes}
and~\ref{tab:materiapps-hnsw}, is not publicly released at present, but is
available from the corresponding author on reasonable request.

\bibliographystyle{tfnlm}
\bibliography{main}

@misc{materiapps,
  title        = {{MateriApps}: A Portal Site of Materials Science Simulation},
  author       = {{MateriApps}},
  howpublished = {\url{https://ma.issp.u-tokyo.ac.jp/}},
  note         = {Accessed 31 July 2026},
  year         = {2026},
}

@misc{repocatalog,
  title        = {Research Software Catalog},
  howpublished = {\url{https://k-yoshimi.github.io/repo-catalog/en/}},
  note         = {Generated catalog site. Accessed 31 July 2026},
  year         = {2026},
}

@misc{cataloguedata,
  title        = {Snapshot Data for ``Building a Research-Software Catalog with a Coding Agent: From Hackathon Prototype to Public Deployment''},
  author       = {Yoshimi, Kazuyoshi},
  howpublished = {ISSP Data Repository, The University of Tokyo},
  note         = {Dataset, version 1.0.0: per-record export and Figure~4 values of the 1 September 2026 catalog build, and the MateriApps retrieval corpus (crawl completed 3 August 2026). \url{https://isspns-gitlab.issp.u-tokyo.ac.jp/k-yoshimi/rsc-agent}},
  year         = {2026},
}

@misc{pasumscatalog,
  title        = {{ISSP PASUMS} Catalog},
  author       = {{Project for Advancement of Software Usability in Materials Science (PASUMS), Institute for Solid State Physics, The University of Tokyo}},
  howpublished = {\url{https://issp-center-dev.github.io/software-catalog/en/}},
  note         = {Institutional deployment of the catalog generator. Accessed 31 July 2026},
  year         = {2026},
}

@article{Yoshimi31122025,
author = {Kazuyoshi Yoshimi and Yuichi Motoyama and Tatsumi Aoyama and Mitsuaki Kawamura and Naoki Kawashima},
title = {Project for advancement of software usability in materials science},
journal = {Science and Technology of Advanced Materials: Methods},
volume = {5},
number = {1},
pages = {2564055},
year = {2025},
publisher = {Taylor \& Francis},
doi = {10.1080/27660400.2025.2564055},
URL = {
https://doi.org/10.1080/27660400.2025.2564055
},
eprint = {
https://doi.org/10.1080/27660400.2025.2564055
}
}

@article{Misawa2025,
author = {Takahiro Misawa and Ai Koizumi and Ryo Tamura and Kazuyoshi Yoshimi},
title = {Exploring utilization of generative {AI} for research and education in data-driven materials science},
journal = {Science and Technology of Advanced Materials: Methods},
volume = {5},
number = {1},
pages = {2535956},
year = {2025},
publisher = {Taylor \& Francis},
doi = {10.1080/27660400.2025.2535956},
url = {https://doi.org/10.1080/27660400.2025.2535956}
}

@misc{Chen2021,
  author       = {Mark Chen and Jerry Tworek and Heewoo Jun and others},
  title        = {Evaluating Large Language Models Trained on Code},
  howpublished = {arXiv:2107.03374},
  year         = {2021},
  doi          = {10.48550/arXiv.2107.03374},
  url          = {https://arxiv.org/abs/2107.03374},
}

@misc{copilot,
  author       = {{GitHub, Inc.}},
  title        = {{GitHub Copilot}},
  howpublished = {\url{https://github.com/features/copilot}},
  note         = {Accessed 31 July 2026},
  year         = {2026},
}

@misc{Yetistiren2023,
  author       = {Burak Yeti{\c{s}}tiren and I{\c{s}}{\i}k {\"O}zsoy and Miray Ayerdem and Eray T{\"u}z{\"u}n},
  title        = {Evaluating the Code Quality of {AI}-Assisted Code Generation Tools:
                  An Empirical Study on {GitHub Copilot}, {Amazon CodeWhisperer}, and {ChatGPT}},
  howpublished = {arXiv:2304.10778},
  year         = {2023},
  doi          = {10.48550/arXiv.2304.10778},
  url          = {https://arxiv.org/abs/2304.10778},
}

@inproceedings{Pearce2022,
  author    = {Hammond Pearce and Baleegh Ahmad and Benjamin Tan and Brendan Dolan-Gavitt and Ramesh Karri},
  title     = {Asleep at the Keyboard? {A}ssessing the Security of {GitHub Copilot}'s Code Contributions},
  booktitle = {2022 {IEEE} Symposium on Security and Privacy ({SP})},
  pages     = {754--768},
  year      = {2022},
  doi       = {10.1109/SP46214.2022.9833571},
  url       = {https://doi.org/10.1109/SP46214.2022.9833571},
}

@inproceedings{Zhang2023,
  author    = {Beiqi Zhang and Peng Liang and Xiyu Zhou and Aakash Ahmad and Muhammad Waseem},
  title     = {Practices and Challenges of Using {GitHub Copilot}: An Empirical Study},
  booktitle = {Proceedings of the 35th International Conference on Software Engineering
               and Knowledge Engineering ({SEKE})},
  pages     = {124--129},
  year      = {2023},
  doi       = {10.18293/SEKE2023-077},
  url = {https://doi.org/10.18293/SEKE2023-077},
}

@inproceedings{Lewis2020,
  author    = {Patrick Lewis and Ethan Perez and Aleksandra Piktus and others},
  title     = {Retrieval-Augmented Generation for Knowledge-Intensive {NLP} Tasks},
  booktitle = {Advances in Neural Information Processing Systems},
  volume    = {33},
  pages     = {9459--9474},
  publisher = {Curran Associates, Inc.},
  year      = {2020},
  url       = {https://proceedings.neurips.cc/paper_files/paper/2020/file/6b493230205f780e1bc26945df7481e5-Paper.pdf},
}

@misc{Wang2024,
  author       = {Liang Wang and Nan Yang and Xiaolong Huang and others},
  title        = {Multilingual {E5} Text Embeddings: A Technical Report},
  howpublished = {arXiv:2402.05672},
  year         = {2024},
  doi          = {10.48550/arXiv.2402.05672},
  url          = {https://arxiv.org/abs/2402.05672},
}

@misc{Chen2024,
  author       = {Jianlv Chen and Shitao Xiao and Peitian Zhang and others},
  title        = {{M3-Embedding}: Multi-Linguality, Multi-Functionality, Multi-Granularity
                  Text Embeddings Through Self-Knowledge Distillation},
  howpublished = {arXiv:2402.03216},
  year         = {2024},
  doi          = {10.48550/arXiv.2402.03216},
  url          = {https://arxiv.org/abs/2402.03216},
}

@misc{Nogueira2019,
  author       = {Rodrigo Nogueira and Kyunghyun Cho},
  title        = {Passage Re-ranking with {BERT}},
  howpublished = {arXiv:1901.04085},
  year         = {2019},
  doi          = {10.48550/arXiv.1901.04085},
  url          = {https://arxiv.org/abs/1901.04085},
}

@article{Malkov2020,
  author  = {Yu A. Malkov and D. A. Yashunin},
  title   = {Efficient and Robust Approximate Nearest Neighbor Search Using
             Hierarchical Navigable Small World Graphs},
  journal = {IEEE Transactions on Pattern Analysis and Machine Intelligence},
  volume  = {42},
  number  = {4},
  pages   = {824--836},
  year    = {2020},
  doi     = {10.1109/TPAMI.2018.2889473},
}

@article{Ji2023,
  author  = {Ziwei Ji and Nayeon Lee and Rita Frieske and others},
  title   = {Survey of Hallucination in Natural Language Generation},
  journal = {ACM Computing Surveys},
  volume  = {55},
  number  = {12},
  pages   = {1--38},
  year    = {2023},
  doi     = {10.1145/3571730},
}

@article{Barker2022,
  author  = {Michelle Barker and Neil P. {Chue Hong} and Daniel S. Katz and others},
  title   = {Introducing the {FAIR} Principles for research software},
  journal = {Scientific Data},
  volume  = {9},
  number  = {1},
  pages   = {622},
  year    = {2022},
  doi     = {10.1038/s41597-022-01710-x},
}

@article{Smith2016,
  author  = {Arfon M. Smith and Daniel S. Katz and Kyle E. Niemeyer and
             {FORCE11 Software Citation Working Group}},
  title   = {Software citation principles},
  journal = {PeerJ Computer Science},
  volume  = {2},
  pages   = {e86},
  year    = {2016},
  doi     = {10.7717/peerj-cs.86},
}

@misc{claudecode,
  author       = {{Anthropic}},
  title        = {{Claude Code}},
  howpublished = {\url{https://claude.com/product/claude-code}},
  note         = {Command-line coding agent. Accessed 31 July 2026},
  year         = {2026},
}

@misc{ollama,
  author       = {{Ollama}},
  title        = {{Ollama}},
  howpublished = {\url{https://ollama.com/}},
  note         = {Local language-model runtime. Accessed 31 July 2026},
  year         = {2026},
}

@article{Ison2016,
  author  = {Jon Ison and Kristoffer Rapacki and Herv{\'e} M{\'e}nager and others},
  title   = {Tools and data services registry: a community effort to document
             bioinformatics resources},
  journal = {Nucleic Acids Research},
  volume  = {44},
  number  = {D1},
  pages   = {D38--D47},
  year    = {2016},
  doi     = {10.1093/nar/gkv1116},
}

@inproceedings{DiCosmo2017,
  author    = {Roberto {Di Cosmo} and Stefano Zacchiroli},
  title     = {Software Heritage: Why and How to Preserve Software Source Code},
  booktitle = {iPRES 2017: 14th International Conference on Digital Preservation},
  address   = {Kyoto, Japan},
  pages     = {1--10},
  year      = {2017},
  url       = {https://hal.science/hal-01590958},
}

@inproceedings{Sculley2015,
  author    = {D. Sculley and Gary Holt and Daniel Golovin and others},
  title     = {Hidden Technical Debt in Machine Learning Systems},
  booktitle = {Advances in Neural Information Processing Systems},
  volume    = {28},
  pages     = {2503--2511},
  publisher = {Curran Associates, Inc.},
  year      = {2015},
  url       = {https://papers.nips.cc/paper/5656-hidden-technical-debt-in-machine-learning-systems},
}

@inproceedings{Sambasivan2021,
  author    = {Nithya Sambasivan and Shivani Kapania and Hannah Highfill and
               Diana Akrong and Praveen Paritosh and Lora M. Aroyo},
  title     = {``Everyone wants to do the model work, not the data work'':
               Data Cascades in High-Stakes {AI}},
  booktitle = {Proceedings of the 2021 {CHI} Conference on Human Factors in
               Computing Systems},
  articleno = {39},
  pages     = {1--15},
  year      = {2021},
  doi       = {10.1145/3411764.3445518},
  url = {https://doi.org/10.1145/3411764.3445518},
}

@inproceedings{Zheng2023,
  author    = {Lianmin Zheng and Wei-Lin Chiang and Ying Sheng and others},
  title     = {Judging {LLM}-as-a-Judge with {MT-Bench} and {Chatbot Arena}},
  booktitle = {Advances in Neural Information Processing Systems:
               Datasets and Benchmarks Track},
  volume    = {36},
  year      = {2023},
  url       = {https://arxiv.org/abs/2306.05685},
}

@misc{Gao2023,
  author       = {Yunfan Gao and Yun Xiong and Xinyu Gao and others},
  title        = {Retrieval-Augmented Generation for Large Language Models: A Survey},
  howpublished = {arXiv:2312.10997},
  year         = {2023},
  doi          = {10.48550/arXiv.2312.10997},
  url          = {https://arxiv.org/abs/2312.10997},
}

@misc{wcag22,
  author       = {{World Wide Web Consortium (W3C)}},
  title        = {Web Content Accessibility Guidelines ({WCAG}) 2.2},
  howpublished = {\url{https://www.w3.org/TR/WCAG22/}},
  note         = {W3C Recommendation, 12 December 2024. Accessed 19 August 2026},
  year         = {2024},
}

@misc{matdacs,
  title        = {{MatDaCs} ({MaterialsDataCommons}): A Portal Site Supporting Data-Driven Research},
  author       = {{Data Creation and Utilization-Type Materials Research and Development Project (DxMT)}},
  howpublished = {\url{https://mat-dacs.dxmt.nims.go.jp/en/}},
  note         = {Accessed 19 August 2026},
  year         = {2026},
}

@misc{aimhack2026,
  title        = {{DxMT AIMHack 2026}: Construction of Materials Databases Using {AI} Agents},
  author       = {{Data Creation and Utilization-Type Materials Research and Development Project (DxMT)}},
  howpublished = {\url{https://dxmt.nims.go.jp/news/4090}},
  note         = {Held 24--26 June 2026, Gotemba, Shizuoka, Japan. In Japanese. Accessed 31 July 2026},
  year         = {2026},
}

@article{Konishi15,
author = {Yusuke Konishi and Ryo Igarashi and Shusuke Kasamatsu and Takeo Kato and Naoki Kawashima and Tsutomu Kawatsu and Hikaru Kouta and Masashi Noda and Shoichi Sasaki and Yayoi Terada and Synge Todo and Shigehiro Tsuchida and Kazuyoshi Yoshimi and Kanako Yoshizawa},
title = {{MateriApps} -- a Portal Site of Materials Science Simulation},
journal = {Proceedings of Computational Science Workshop 2014 (CSW2014), {JPS} Conf. Proc.},
volume = {5},
pages = {011007},
year = {2015},
doi = {10.7566/JPSCP.5.011007},
url = {https://journals.jps.jp/doi/abs/10.7566/JPSCP.5.011007}
}

@article{MAL-paper,
author = {Yuichi Motoyama and Kazuyoshi Yoshimi and Takeo Kato and Synge Todo},
title = {{MateriApps} {LIVE}! and {MateriApps Installer}: Environment for starting and scaling up materials science simulations},
journal = {{SoftwareX}},
volume = {20},
pages = {101210},
year = {2022},
publisher = {Elsevier {BV}},
doi = {10.1016/j.softx.2022.101210},
url = {https://doi.org/10.1016/j.softx.2022.101210}
}

@misc{openaicodex,
  author       = {{OpenAI}},
  title        = {{Codex CLI}},
  howpublished = {\url{https://developers.openai.com/codex/cli}},
  note         = {Command-line coding agent. Accessed 19 August 2026},
  year         = {2026},
}

@inproceedings{hanai2024arim,
    title = {{ARIM-mdx Data System}: Towards a Nationwide Data Platform for Materials Science},
    author = {Hanai, Masatoshi and Ishikawa, Ryo and Kawamura, Mitsuaki and Ohnishi, Masato and Takenaka, Norio and Nakamura, Kou and Matsumura, Daiju and 
        Fujikawa, Seiji and Sakamoto, Hiroki and Ochiai, Yukinori and Okane, Tetsuo and Kuroki, Shin-Ichiro and Yamada, Atsuo and Suzumura, Toyotaro and
        Shiomi, Junichiro and Tarua, Kenjiro and Mita, Yoshio and Shibata, Naoya and Ikuhara, Yuichi},
    booktitle = {Proceedings of 2024 IEEE International Conference on Big Data (BigData)},
    pages={2326-2333},
    year = {2024},
    doi = {10.1109/BigData62323.2024.10825674},
    url = {https://doi.org/10.1109/BigData62323.2024.10825674},
}

\end{document}